\documentstyle[12pt]{article}
\begin{document}

\begin{center}
{\large \bf Pulsed pump in optical displacement transducer 
for experiments with probe bodies} \\ [3mm]

Victor V. Kulagin\\ [2mm]

{\it Sternberg Astronomical Institute, Moscow State 
University,\\
 Universitetsky prospect 13, 119899, Moscow, Russia, 
e-mail: kul@sai.msu.ru}\\ [5mm]

\end{center}

\begin{abstract}
The sensitivity of the displacement transducer pumped with a
train of high-intensity laser pulses is estimated. Due to the
multicomponent character of the pump a consideration of
transformations of the signal and the noises between optical modes
plays an important role in estimation of the potential sensitivity.
An expression for the minimal detectable external classical force
resembles those for the continuous wave pumping with substitution of
the laser power by a time averaged power of pulsed laser. Possible
scheme for back action noise compensation for such transducers is
considered. For full suppression of back action noise the field of 
local oscillator has to be pulsed with the same time dependence as 
the pump field.

\end{abstract}


\section{Introduction}

\par
The longbase laser interferometric gravitational wave detectors
are under construction at present time [1-3]. Their sensitivity to
metric perturbation will be about $h \approx  10^{-21}$ that corresponds to 
the
classical regime of operation. However for future installations with
projected sensitivity $10^{-22} \div  10^{-23}$ the quantum features of the
measurement process can play a significant role. At the same time
there are no limits of principle on the accuracy of measurement of
external classical force. Therefore the methods and schemes which
give the possibility to overcome the quantum measurement limitations
(or the so called standard quantum limit, SQL) is of vital
importance for future generation of gravitational wave experiments.
\par
There are several procedures which allow to achieve the
sensitivity larger than the SQL [4,5]. For example in [5] an optimal
filtration procedure for the simplest variant of the optical sensor
- a mirror attached to a mechanical resonator and illuminated with a
coherent pump field - was considered. An external force acting on
the mechanical oscillator displaces its equilibrium position and
thus changes the phase of the reflected field. The vacuum
fluctuations of the input light act on the oscillator through the
radiation pressure effect and constitute the back action noise of
the measuring apparatus. For such system two quadratures of
reflected wave are correlated. Using correlation (phase sensitive)
processing of two quadratures one can increase a signal-to-noise
ratio and overcome the SQL.
\par
However the gain in sensitivity for the schemes overcoming the SQL is 
usually proportional to the
square root of the ratio of laser power used for pumping the
interferometer and an optimal power that corresponds to the point
where the sensitivity of the interferometer achieve the SQL [4-6].
Unfortunately the optimal power is impracticably large, about
several dozens of kilowatts that restrains the experimental
implementation of the technique.
\par
The pumping with the ultrashort periodic laser pulses can be 
technically advantageous over a continuous wave pumping for 
practical realization of the schemes overcoming the SQL. Actually 
for a large power a problem of generating a train of short high-intensity 
laser pulses can be technically easier than a problem of 
cw light generation (when the averaged powers for two cases are 
 equal) because in the first case 
the energy in laser resonator is spread over the large frequency 
band (and different spatial longitudinal modes) and high intensities 
can be produced relatively easily. At the same time the amplitude 
and frequency stability of the pulsed pump in the case of a mode 
locked laser can be at the same level as for the monochromatic pump 
[7,8]. For example in [8] the stability of intermode beats for 
the mode locked laser output was estimated as $5\cdot 
10^{-12}$  in 10 s.

Another consideration is that the perspectives of squeezed states
generation with high nonclassicality seem more realistic 
for the case of short high-intensity laser
pulses allowing the use of squeezed pulsed pump
in displacement transducers [9].

     Finally an analog to digital conversion 
is usually used in modern experiment during the processing of the 
output. Therefore it seems natural to take the pulsed pump at once 
so that the output will comprise a set  of the values for appropriate 
variable at 
definite times. 
\par
The goals of this article are to consider a displacement
transducer consisting of a mirror attached to a mechanical
oscillator and illuminated with a train of high-intensity laser
pulses, to reveal the algorithm of optimal signal processing for
such transducer and to estimate the sensitivity of the scheme to a
measurement of classical external force.
\par
The model of displacement transducer and basic equations of
motion is considered in section 2. The sensitivities for traditional
measurement scheme and for correlative processing of the output
quadratures in the case of time independent pump are estimated in
sections 3 and 4 correspondingly. The pulsed pump for the
displacement transducer is considered in section 5. The conclusions are
in section 6.
\par
\section{ Model for displacement transducer and transformation of
quadrature components}
\par
Let consider the most simple case of optical displacement
transducer - a mirror attached to a mass of a mechanical oscillator
and illuminated with a train of high-intensity laser pulses. An
external force displaces an equilibrium position of mechanical
oscillator changing the phase of reflected wave. The variation of
the reflected field phase is measured by a readout system. This
model is easy to calculate and it contains at the same time all
features of displacement transducers with pulsed pump. For the
incident $E_{\rm i}$ and reflected $E_{\rm r}$ waves one can use the
quasimonochromatic approximation
\par
\begin{eqnarray}
E_{\rm i} & = & (A(t-x/c)+a_{1})\cdot \cos \omega _{\rm p}(t-x/c) - 
a_{2}\cdot \sin \omega _{\rm p}(t-x/c)\nonumber \\
E_{\rm r} & = & (B(t+x/c)+b_{1})\cdot \cos \omega _{\rm p}(t+x/c) - 
b_{2}\cdot \sin \omega _{\rm p}(t+x/c)   \label{1}
\end{eqnarray}
\noindent where $A(t-x/c)$ and $\omega _{\rm p}$ are an amplitude 
(mean value) and 
a frequency
of the pump wave, $a_{1}$ and $a_{2}$ are the operators of the quadrature
components (fluctuations) of the pump wave (vacuum for coherent
state$), B(t+x/c)$ is an amplitude (mean value) of the reflected wave,
$b_{1}$ and $b_{2}$ are the operators of the quadrature components
(fluctuations) of the reflected wave. The periodic envelope function
$A(t-x/c)$ consists of a train of equally spaced pulses and the
duration of each pulse is considerably larger than the period of
light wave but considerably smaller than the period of the
mechanical oscillator.
\par
To obtain the equation coupling the amplitudes of the incident
and reflected waves for the moving mirror one can use a
transformation of electromagnetic field for moving reference frame
[10]. For a constant velocity of the mirror $V$ one has
\par
\begin{equation}
E_{\rm r} = -[(1-V/c)/(1+V/c)]\cdot E_{\rm i}\exp (-2{\rm i} \omega 
_{\rm p}X/c)
\label{2}\end{equation}
\noindent where for simplicity the reflection coefficient of the mirror is
taken to be $r \approx  -1$ and $X$ is the position of the mirror. Let suppose
that this expression is valid also for the slowly varying velocity
$V(t)$ and position $X(t)$ of the mirror and $\mid V(t)\mid  \ll  c$ (the 
validity of
equation (2) has been proved for the mirror consisting of free
electrons for the general case of relativistic velocity $V(t)$ in 
[11]). Then in linear approximation in $V/c$ one can obtain from 
equation
(2) the following expression
\par
\begin{equation}
E_{\rm r} = -(1 - 2V(t)/c - 2{\rm i} \omega _{\rm p}X(t)/c)\cdot E_{\rm i}
\label{3}\end{equation}
The first term in (3) is an amplitude modulation of the
reflected wave due to the mirror movement and the second is a phase
modulation. For slow motion of the mirror $V \approx   \omega  _{\mu 
}X$ ($\omega _{\mu }$ is a frequency
of mechanical oscillator) and the second term in brackets is
considerably smaller than the third term. Therefore for the
transformation of the quadrature components of the field one can
obtain
\par

\begin{eqnarray}
b_{1}(t) & = & - a_{1}(t)
\nonumber \\
b_{2}(t) & = & - a_{2}(t) + 2\omega _{\rm p}A(t)X(t)/c
\label{4}
\end{eqnarray}
For the equation of mirror motion one has
\par
\begin{equation}
\ddot X(t) + 2\delta _{\mu }\dot X(t) + \omega ^{2}_{\mu }X(t) = 
M^{-1}(F_{\rm s}(t) + F_{\rm p}(t) + F_{\rm th}(t))
\label{5}\end{equation}
\noindent where $M$ and $\delta _{\mu }$ are the mass and the damping 
coefficient of
mechanical oscillator, $F_{\rm s}(t)$ is a signal force, $F_{\rm p}(t)$ 
is radiation
pressure force and $F_{\rm th}(t)$ is a force associated with the damping 
of
the oscillator. Let suppose for simplisity that $\delta _{\mu }$ tends to 
zero.
Then the displacement $X(t)$ of the mirror will consist of two parts -
a signal displacement $X_{\rm s}(t)$ and a radiation pressure displacement
$X_{\rm p}(t)$. For $F_{\rm p}(t)$ one has
\par
\begin{equation}
F_{\rm p}(t) = SA(t)\cdot a_{1}(t)/(4\pi )
\label{6}\end{equation}
\noindent where $S$ is a cross section of the laser beam. Therefore the
equations of motion for the displacement transducer are
\par
\begin{eqnarray}
b_{1}(t) & = & - a_{1}(t)
\nonumber \\
b_{2}(t) & = & - a_{2}(t) + 2\omega _{\rm p}A(t)X(t)/c
\label{7} \\
\ddot X(t) & + & 2\delta _{\mu }\dot X(t) + \omega ^{2}_{\mu }X(t) = 
M^{-1}(F_{\rm s}(t) + SA(t)\cdot a_{1}(t)/(4\pi )) \nonumber 
\end{eqnarray}

\section{ Sensitivity for a traditional measurement scheme}
\par
For traditional measurement scheme [4,6] the amplitude of the
pump is constant. Therefore one can easily obtain the transformation
relations for the quadratures $b_{1}$ and $b_{2}$ from equations (7)
\par
\begin{eqnarray}
b_{1}(\omega ) & = & - a_{1}(\omega )
\nonumber \\
b_{2}(\omega ) & = & - a_{2}(\omega ) + \lambda \xi (\omega 
)A^{2}a_{1}(\omega ) + A\xi (\omega )F_{\rm s}(\omega )
\label{8}
\end{eqnarray}
\noindent where $\xi (\omega ) = 2\omega _{\rm p}G(\omega )/c$, $G(\omega ) = 
\left[M(-\omega ^{2} - 2\delta _{\mu }{\rm i} \omega + \omega ^{2}_{\mu 
})\right]^{-1}$ is mechanical
oscillator transfer function and $\lambda  = S/(4\pi )$.
\par
Only quadrature $b_{2}$ contains the signal and it is this
quadrature that is measured in traditional measurement scheme [4,6].
This corresponds to the measurement of the phase of the reflected
wave. The first term in the right hand side of equation (8) for $b_{2}$
can be treated as an additive noise and the second term as a back
action noise. For small pump amplitudes the sensitivity is
increasing with the increase of $A$ because the signal is proportional
to $A$. However for large pump amplitudes the second term in r.h.s. of
(8) becomes dominant and the sensitivity is decreasing with the
increase of $A$. Therefore there is an optimal value of the pump
amplitude and the sensitivity to external force at this pump
amplitude is just the SQL [6].
\par
\section{Correlative processing of quadratures for time independent
amplitude of the pump}
\par
Two quadratures of the reflected field according to equation
(8) have the dependence on the amplitude fluctuations of the
incident field $a_{1}$. Therefore one can expect that the sensitivity can
be increased for the correlative processing of the output [5,12].
Actually if one combine with appropriate weight coefficients the
quadratures $b_{1}$ and $b_{2}$ of the output wave then in this combination
the noise term depending on $a_{1}$ can be cancelled. This weighting can
be done by a homodyne detector with appropriate choise of a local
oscillator phase $\phi $.
\par
Let the field of the local oscillator have the form
\par
\begin{equation}
E_{\rm L}(t) = A_{\rm L}\cos (\omega _{\rm p}t+\phi )
\label{9}
\end{equation}
\noindent Then the photodetector output is proportional to the following
expression according to (1), (9)
\par
\begin{equation}
I_{\rm pd} \propto  A_{\rm L}(b_{1}\cos \phi  + b_{2}\sin \phi )
\label{10}
\end{equation}
\noindent and at certain frequency $\omega _{\rm f}$ one can obtain
\par
\begin{equation}
 I_{\rm pd} \propto  A_{\rm L}[a_{1}(\omega _{\rm f})(-\cos \phi +\lambda 
\xi (\omega _{\rm f})A^{2}\sin \phi ) - a_{2}(\omega  _{\rm  f})\sin 
\phi + A\xi (\omega _{\rm f})F_{\rm s}(\omega _{\rm f})\sin \phi ]
\label{11}
\end{equation}
\noindent Therefore choosing the phase  $\phi  $  according  to  the 
equation ($\xi (\omega _{\rm f})$ is
real for $\delta _{\mu } = 0$)
\par
\begin{equation}
-\cos \phi  + \lambda \xi (\omega _{\rm f})A^{2}\sin \phi  = 0
\label{12}
\end{equation}
\noindent one can make the photocurrent insensitive to the amplitude
fluctuations $a_{1}$ of the input field at certain frequency $\omega 
_{\rm f}$ of the
signal. In this case the increase of the pump amplitude $A$ results in
the relative increase of the output signal at frequency 
$\omega _{\rm f}$ according to the equation
(11) with respect to the noise level defined by $a_{2}$. 
\par
For compensation of the back action noise inside definite
frequency band one has to use the time dependent local oscillator
phase $\phi (t)$ [12,13]. In this case the optimal dependence of 
$\phi $ on $t$ is
defined by the displacement transducer transfer function 
$\xi (\omega _{\rm f})$ and by the
spectrum of the external force $F_{\rm s}(\omega )$ [12].

So a signal-to-noise ratio is proportional to $A^{2}$ (there is no 
optimal
power) and in principle there is no sensitivity limitation by the 
SQL.
In real experiment when the pump power gets larger the output signal and
noises get smaller according to equation (11) if the condition (12)
is kept valid therefore when $A$ becomes greater than 
a certain value then the noises of
photodetector electronics can limit the sensitivity. However this
noises have technical character and will be neglected in the
following.

Another sensitivity restriction can arise due to the damping in 
mechanical oscillator (mirror) [14,15]. This problem is general for 
all supersensitive measurements. At the same time an intrinsic 
dissipation obtained in modern experiments for mechanical oscillator 
is far larger (by several orders of magnitude) than the value 
expected from the first principles [16] therefore it can be treated 
also as a technical problem now and will not be adressed below.

It is worth to mention that the increase in sensitivity over the 
usual measurement scheme occurs here due to utilization of the 
internal squeezing 
(self-squeezing) of the reflected beam because of the nonlinear 
(quadratic) interaction of the incident light and the mirror 
[17,18]. Actually two quadratures of the reflected beam are 
correlated and it is this fact that allow to use the correlative 
processing of the output. On the other hand the correlation of 
the quadratures according to equations (8) means the squeezing of 
the beam and the larger the correlation coefficient 
$\lambda \xi (\omega )A^{2}$ the larger the internal squeezing [17].

\par
\section{Sensitivity for the pulsed pump}
\par
Let consider the periodic envelope $A(t)$ which consists of a
train of equally spaced pulses with duration $\tau $ and period $T$. The
spectrum of this pump has also the form of a train of pulses in
frequency domain with the distance between neighbour pulses
\par
\begin{equation}
\omega _{\rm q} = 2\pi T^{-1}
\label{13}
\end{equation}
For the amplitude of the pump $A(t)$ one can use now the
expansion into the Fourier series
\par
\begin{equation}
A(t) = \sum^{\infty }_{n=-\infty } g_{n}\exp (-{\rm i} n\omega _{\rm q}t)
\label{14}
\end{equation}
\par
\noindent and the particular form of $A(t)$ is defined by the set of Fourier
amplitudes $g_{n}$.
\par
The response of the displacement transducer now have many
frequency components at $\omega  = n\omega _{\rm q}, n = 0, 1 \ldots 
$ according to the
equations (4) and each frequency component contains the signal part
besides the radiation pressure force $F_{\rm p}(t)$ have also wide spectrum
(cf. (6)). So there are two problems: how to collect the signal
parts from the whole spectral band of the output and how to achieve
the compensation of the radiation pressure noise in the output. 
It is clear that the monochromatic local oscillator is inappropriate 
for the homodyning because quadrature 
$b_{1}(t)$ of the output signal contains in this case 
the quadrature $a_{1}(t)$ of the input noises only from one 
frequency and the radiation pressure force $F_{\rm p}(t)$ in 
expression for $b_{2}(t)$ (cf. (7)) contains 
$a_{1}(t)$ from all frequencies $n\omega _{\rm q}$ therefore the 
full compensation is impossible.

Fortunately two problems can be overcome by the use of the pulsed
local oscillator with the amplitude time dependence resembling that
for the pump.
\par
For the radiation pressure displacement $X_{\rm p}$ of the mechanical
oscillator one has from equations (5), (6) and (14) the following
expression
\par
\begin{equation}
X_{\rm p}(\omega ) = G(\omega )F_{\rm p}(\omega ) = \lambda G(\omega 
)\sum^{\infty 
}_{n=-\infty } g_{n}a_{1}(\omega -n\omega _{\rm q})
\label{15}
\end{equation}
For the quadrature transformation one can obtain instead of (8)
the following equations from (4) and (14)
\par
\begin{eqnarray}
b_{1}(\omega ) & = & - a_{1}(\omega ) \nonumber \\
b_{2}(\omega  )  &  =  &  -   a_{2}(\omega   )   +   2\omega   _{\rm 
p}c^{-1}\sum^{\infty 
}_{k=-\infty } g_{k}(X_{\rm p}(\omega -k\omega _{\rm q}) + 
X_{\rm s}(\omega -k\omega _{\rm q})) \label{16}
\end{eqnarray}
\par
Let suppose the local oscillator field in the form of
\par
\begin{equation}
E_{\rm L}(t) = A_{\rm L}(t)\cos (\omega _{\rm p}t+\phi )
\label{17}
\end{equation}
\par
\noindent where the dependence of the amplitude $A_{\rm L}(t)$ on $t$ is much 
slower than
$\cos \omega  _{\rm  p}$t.  Then  for  the  envelope  of  the  local 
oscillator field $A_{\rm L}(t)$ the
Fourier expansion similar to (14) is valid
\par
\begin{equation}
A_{\rm L}(t) = \sum^{\infty }_{n=-\infty } e_{n}\exp (-{\rm i} n\omega 
_{\rm q}t)
\label{18}
\end{equation}
The photodetector current has now the following form
\par
\begin{equation}
I_{\rm pd} \propto   A_{\rm L}(t)(b_{1}(t)\cos \phi  + b_{2}(t)\sin \phi )
\label{19}
\end{equation}
\noindent and in the frequency domain one has
\par
\begin{equation}
I_{\rm pd}(\omega ) \propto  \cos \phi \cdot \sum^{\infty }_{n=-\infty 
}e_{n}b_{1}(\omega -n\omega _{\rm q}) + \sin \phi \cdot \sum^{\infty 
}_{n=-\infty }e_{n}b_{2}(\omega -n\omega _{\rm q})
\label{20}
\end{equation}
Let consider different parts in the photodetector output. The
first term in equation (20) depends only on the amplitude
fluctuations of the input field according to (16)
\par
\begin{equation}
\cos \phi \cdot \sum^{\infty }_{n=-\infty }e_{n}b_{1}(\omega -n\omega _{
\rm q}) = 
- \cos \phi \cdot \sum^{\infty }_{n=-\infty }e_{n}a_{1}(\omega -n\omega _{
\rm q})
\label{21}
\end{equation}
The second term in equation (20) contains the signal and the
noise parts. The noise part consists of the additive noise and the
back action noise and has the following expression according to (15)
and (16)
\par
\begin{eqnarray}
 & & [\sin \phi \cdot \sum^{\infty }_{n=-\infty }e_{n}b_{2}(\omega -n\omega 
_{\rm q})]_{\rm noise} = - \sin \phi \cdot \sum^{\infty }_{n=-\infty 
}e_{n}a_{2}(\omega -n\omega _{\rm q}) + 
2\omega _{\rm p}c^{-1}\sin \phi \cdot \lambda \cdot \nonumber \\
 & & \sum^{\infty }_{n=-\infty 
}\sum^{\infty }_{k=-\infty }e_{n}g_{k}G(\omega -k\omega _{\rm q}-n\omega 
_{\rm q})\{\sum^{\infty }_{m=-\infty }g_{m}a_{1}(\omega -k\omega _{
\rm q}-n\omega _{\rm q}-m\omega _{\rm q})\}
\label{22}
\end{eqnarray}
Let consider only the photocurrent at small frequencies $\omega  \approx  
\omega _{\mu }$.
Then the main input into the photocurrent will be given by the
resonant terms for which $k+n=0$. With this supposition one has from
equation (22)
\par
\begin{eqnarray}
& & [\sin \phi \cdot \sum^{\infty }_{n=-\infty }e_{n}b_{2}(\omega -n\omega 
_{q})]_{\rm noise} = - \sin \phi \cdot \sum^{\infty }_{n=-\infty 
}e_{n}a_{2}(\omega -n\omega _{\rm q}) + \nonumber \\
& & \sin \phi \cdot \lambda \xi (\omega )\cdot \sum^{\infty }_{m=-\infty 
}e_{m}g_{-m}\cdot \sum^{\infty }_{n=-\infty }g_{n}a_{1}(\omega -n\omega 
_{\rm q})\}
\label{23}
\end{eqnarray}

Comparing equations (21) and (23) one can conclude that full
compensation of back action noise is possible only for
\par
\begin{equation}
e_{n} = \alpha g_{n}
\label{24}
\end{equation}
where $\alpha $ is the same for all numbers $n$ so the forms of pump and local
oscillator fields have to be the same (apart from the scale factor
$\alpha $).
\par
Let now consider the signal part of the second term in the
r.h.s. of equation (20). From equations (7), (16) and (20) one has
\par
\begin{eqnarray}
& & [\sin \phi \cdot \sum^{\infty }_{n=-\infty }e_{n}b_{2}(\omega -n\omega 
_{\rm q})]_{\rm signal} = \nonumber \\
& & \sin \phi \cdot \sum^{\infty }_{n=-\infty }e_{n}\{\sum^{\infty 
}_{k=-\infty 
}g_{k}\xi (\omega -k\omega _{\rm q}-n\omega _{\rm q})F_{\rm 
s}(\omega -k\omega _{\rm q}-n\omega _{\rm q})\}
\label{25}
\end{eqnarray}
Evaluation of this expression for the condition $k+n=0$ gives
\par
\begin{equation}
[\sin \phi \cdot \sum^{\infty }_{n=-\infty }e_{n}b_{2}(\omega -n\omega 
_{\rm q})]_{\rm signal} = 
\sin \phi \cdot \xi (\omega )F_{\rm s}(\omega ) 
\sum^{\infty }_{n=-\infty }e_{n}g_{-n}
\label{26}
\end{equation}
Combining equations (20), (23), (24) and (26) and supposing
that the back action noise is compensated in the output of the
photodetector one can obtain for the spectral density of noises in
the photocurrent the following expression
\par
\begin{equation}
N(\omega ) \propto  \sin \phi \cdot N_{0}\cdot \sum^{\infty }_{n=-\infty 
}g_{n}g_{-n} = \sin \phi \cdot N_{0}P
\label{27}
\end{equation}
\noindent where it is supposed that fluctuations at frequencies $\omega  - 
n\omega _{q}, n =0, 1\ldots $ are uncorrelated and have the same 
spectral density $N_{0} $ (this
assumption is valid for not very small duration of pump pulses), $P$
is proportional to the time averaged power of the pulsed pump. Then
for the signal-to-noise ratio $\mu $ one has from equations (26) and (27)
the following expression
\par
\begin{equation}
\mu  \propto  N^{-1}_{0}P\int ^{\infty }_{-\infty }\mid \xi (\omega 
)F_{\rm s}(\omega )\mid ^{2}{\rm d}\omega  
\label{28}
\end{equation}

This value is just equal to the signal-to-noise ratio for 
continuous wave pump with
a power $P$ and correlative processing of the output (cf. equation
(11)). Note that the sensitivity here is not limited by the SQL like
in the case of correlative processing of quadratures for the
monochromatic pump and is increasing with the increase of $P$.
\par
It is worth to mention that the condition for the back action
noise compensation for the pulsed pump is just the same as for the
monochromatic pump (cf. equation (12)) with substitution of the $A^{2}$
with the time averaged value $P$. Therefore the compensation of the
back action noises for the finite frequency band can be possible for
the time varying phase of the local oscillator [12, 13].
\par
\section{Conclusion}
\par

The pumping of the displacement transducer with a train of the
short high-intensity laser pulses is considered. The algorithm of 
optimal signal processing for such transducer is revealed. It 
consists of the correlative processing of the output using the 
pulsed local oscillator with the same envelope as for 
the pump field (apart from the scale factor). In this case the back 
action noise due to the radiation pressure force can be fully 
compensated and the sensitivity of the scheme to a detection of 
a classical external force is not limited by the SQL (as for the case 
of correlative quadrature processing and monochromatic pump field). 

The pulsed pump can be advantageous over the single frequency 
pumping when the nonlinear optical elements are used unside the 
system. Thus considerable increase in sensitivity can be achieved 
for a gravitational interferometric Fabry-Perot type detector with 
a nonlinear optical element placed in a waist of the beam [19]. 
The use of the phase-conjugate mirrors in a gravitational detector of the 
Michelson type allows to construct the system with the parallel arms 
[20]. For such systems an efficiency depends on the instant 
power of the light beam and can be high for the short 
intensive pulses.

\par
In this article only the problem of the force detection with
known spectrum is considered. The reconstruction of unknown external
force acting on the displacement transducer with the pulsed pump
below the standard quantum limit will be considered elsewhere.
\par


\end{document}